\documentclass[12pt]{article}
\usepackage{acronym}
\usepackage{amsmath}
\usepackage{amssymb}
\usepackage{booktabs}
\usepackage{caption}
\usepackage{comment}
\usepackage{graphicx}
\usepackage{longtable}
\usepackage{multicol}
\usepackage{multirow}
\usepackage{placeins}
\usepackage{verbatim}
\usepackage{xcolor}
\usepackage{mathtools, cuted}
\usepackage{vmargin}
\setmarginsrb{15mm}  
             { 5mm}  
             {15mm}  
             {10mm}  
             {10pt}  
             { 5mm}  
             {10pt}  
             {10mm}  

\newcommand{\bs}[1]{\boldsymbol{#1}}
\newcommand{\wt}[1]{\widetilde{#1}}
\newcommand{\mc}[1]{\mathcal{#1}}
\acrodef{2C-2D}{two-component-two-dimensional}
\acrodef{BBB}{BeagleBone Black}
\acrodef{HPOD}{Hilbert proper orthogonal decomposition}
\acrodef{MCCD}{multigrid/multipass cross-correlation digital}
\acrodef{SVD}{singular value decomposition}
\acrodef{POD}{proper orthogonal decomposition}
\acrodef{DMD}{Dynamic mode decomposition}
\acrodef{DMDc}{dynamic mode decomposition with control}
\acrodef{EDMD}{extended dynamic mode decomposition}
\acrodef{OptDMD}{optimised dynamic mode decomposition}
\acrodef{PIV}{particle image velocimetry}
\acrodef{REF}{reference}
\acrodef{RRR}{rank reduction ratio}
\acrodef{PRS}{porous}
\acrodef{SHS}{superhydrophobic}
\acrodef{SP}{sparsity-promoting}
\acrodef{TDMD}{total dynamic mode decomposition}
\acrodef{TKE}{turbulent kinetic energy}
\title{Investigation of the Effects of Superhydrophobic Surface Treatment on the Dynamics of the Flow in the Near Wake of a Sphere using Spatial Dynamic Mode Decomposition} 
\author{Shaun Davey, Callum Atkinson and Julio Soria}
\date{Laboratory for Turbulence Research in Aerospace and Combustion,\\
      Department of Mechanical and Aerospace Engineering,
      \\Monash University, Melbourne, 3800, Victoria, Australia}
\begin{document}
\maketitle
\begin{abstract}
  Viscous drag arises from the fluid at a surface having zero relative velocity, a phenomenon known as the no-slip condition.
  Superhydrophobic surfaces, when submerged in water, trap a layer of air in their surface texture, partially replacing the liquid-solid interface with a liquid-gas interface.
  This air layer, called the plastron, results in partial slip at the surface, thereby reducing the viscous drag.
  In turbulent flows, large fluctuations in pressure and velocity can deplete or completely remove the plastron from the surface. This makes evaluating the effects of superhydrophobic surface treatments on flow dynamics particularly challenging.
  This study examines the impact of a sustained plastron on the dynamics in the shear layer of a sphere, achieved by supplying air at low pressure through pores in the sphere's surface.
  Instantaneous planar velocities in the wakes of spheres, both with and without superhydrophobic surface treatment, are measured within a plane passing through the sphere centres.
  Dynamic mode decomposition (DMD) is applied to the velocity fluctuations in the shear layer to evaluate how superhydrophobic surface treatment affects the instabilities there.
  It is shown that the addition of the pores has a relatively small effect on the instabilities in the shear layer, while they are significantly changed by the addition of superhydrophobic surface treatment when the plastron is sustained.
\end{abstract}

\section{Introduction}
\subsection{Superhydrophobic Surfaces}

The lotus leaf exhibits a unique ability to repel water due to a combination of surface roughness and chemical properties, enabling water droplets to roll off without wetting the surface~\cite{neinhuis1997characterization,samaha2012superhydrophobic}.
This phenomenon, termed superhydrophobicity, has garnered significant interest for its potential applications, particularly in drag reduction and anti-fouling technologies~\cite{rothstein2010slip,bixler2012biofouling}.
Water droplets on a superhydrophobic surface form with a large contact angle, which is the internal angle between the surface and the edge of the droplet~\cite{quere2005non,young1805iii}.
When sliding down an inclined surface, the droplets exhibit low contact angle hysteresis, which is the difference between the angle of the droplet at its leading and trailing edges~\cite{gao2006contact,rothstein2010slip}.
The texture of superhydrophobic surfaces consists of multiscale roughness, which amplifies the effect of chemical hydrophobicity~\cite{wang2020design}.
Due to this combination of properties, the droplets are easily perturbed~\cite{bico1999pearl,chen1999ultrahydrophobic,sakai2006direct,shastry2006directing}.

These properties are achieved by trapping a layer of air, known as the plastron, between the surface and the droplet in what is known as the Cassie-Baxter state~\cite{cassie1944wettability}.
When submerged in water, the air trapped at the surface leads to a partial air-water interface and thus a partial slip over the surface~\cite{rothstein2010slip}, which reduces the skin friction at the surface and decreases the viscous drag of flow over the surface.
A major factor in the varying results of experiments with superhydrophobic surfaces in turbulent flows is the propensity for the plastron to be depleted or completely deteriorated by the large velocity and pressure fluctuations present in turbulent flows.
Even in cases where the plastron remains attached to the surface, these fluctuations have a significant effect on its distribution over the surface~\cite{castagna2018wake}.
This was shown to affect the transverse velocity of a settling sphere by Castagna et al.~\cite{castagna2021onset}.
Once the plastron is depleted, the surface roughness begins to protrude into the flow, and the drag increases.
This can occur across the entire surface as the plastron deteriorates, or in local regions as the plastron is deformed by the forces exerted on it by the flow.

As a plastron-covered superhydrophobic surface decreases viscous drag in water flows, it is of particular interest in reducing fuel consumption in marine vessels, such as ships and submarines, and for reducing transport costs in water supply pipes.
While the drag reduction of superhydrophobic surfaces in laminar flows is directly related to the amount of slip produced, which is determined by the fraction of the surface aerated by the plastron~\cite{rothstein2010slip}, drag reduction in turbulent flows is more complicated to determine. 
While drag reduction has been achieved in turbulent flows, such as boundary layers~\cite{park2014superhydrophobic} and Taylor-Couette flows~\cite{hu2017significant}, investigations of superhydrophobic drag reduction have also resulted in negligible changes~\cite{peguero2009drag}, or even significant increases~\cite{bullee2020bubbly} in drag.
The depletion of the plastron in turbulent boundary layers at high Reynolds number in the study by Aljallis et al.~\cite{aljallis2013experimental} resulted in an increase in drag, while the reduction of drag on a boat in open water observed by Xu et al.~\cite{xu2020superhydrophobic} relied on the plastron being sustained.
The study of turbulent boundary layers on superhydrophobic surfaces by Gose et al.~\cite{gose2018characterization}, over a Reynolds number range of 10,000 to 30,000, resulted in drag effects ranging from a 90\% reduction to a 90\% increase.
This depended on the Reynolds number and surface properties, with the product of the surface-to-projected area ratio of the wetted region and the proportion of the surface area that is aerated being the best predictor of drag reduction.

\subsection{Flow Structures in the Wake of A Sphere}

As the application of superhydrophobic surfaces for drag reduction is of particular interest for maritime vessels, such as ships and submarines, the effects of superhydrophobic surface treatment on the flow over bluff bodies are an important consideration to assess the efficacy of these surfaces.
Spheres serve as an idealised model for axisymmetric bluff bodies, exhibiting reverse flow, flow separation, and periodic and oscillatory vortex shedding in their wakes~\cite{taneda1956experimental,achenbach1972experiments,tomboulides1993direct}.
Quantitative and qualitative measurements in the wake of the sphere have identified numerous structures and characterised their development over various Reynolds numbers.
The wake loses axisymmetry around $Re_D \approx 20$, with a vortex ring forming at $Re_D \approx 25$~\cite{taneda1956experimental,tomboulides2000numerical}.
The vortex ring begins to oscillate at $Re_D \approx 130$, and elongates as the Reynolds number increases.
Axisymmetry in the wake is broken by a pitchfork bifurcation around $Re_D=212$~\cite{frantz2025bifurcation}.
A Hopf bifurcation is formed around $Re_D=275$~\cite{frantz2025bifurcation} after which the flow is dominated by single-frequency vortex shedding~\cite{wu1993sphere,johnson1999flow,tomboulides1993direct}.
A quasiperiodic regime with a second dominant frequency begins at $350 < Re_D < 450$, after the emergence of a Neimark-Sacker bifurcation around $Re_D=330$~\cite{frantz2025bifurcation}.
The generation of a lateral force~\cite{taneda1978visual,yun2006vortical} results in the loss of planar symmetry by $Re_D \approx 500$.
Small-scale vortical structures are present within the main vortex ring from $Re_D \approx 1,000$ as the vortex rings are shed from the sphere and break down as they move downstream~\cite{yun2006vortical,sakamoto1990study}.

The drag of a sphere in motion in a fluid at rest can be determined using settling sphere experiments, in which the sphere is released from rest and accelerates under gravity until it reaches its terminal velocity $w_\infty$.
The corresponding Reynolds number and drag coefficient~\cite{mordant2000velocity} are given by
\begin{equation}\label{eq:Re_inf}
  Re_\infty = \frac{w_\infty d}{\nu}
\end{equation}
and
\begin{equation}\label{eq:CD_inf}
  C_{D,\infty} = \frac{4 d g (\zeta - 1)}{3 w_\infty^2}
\end{equation}
respectively, where $d$ is the sphere diameter, $\nu$ is the kinematic viscosity of water, $g$ is the gravitational acceleration, and $\zeta$ is the sphere-to-fluid density ratio.
In order to measure unsteady drag prior to the sphere reaching terminal velocity, the acceleration of the sphere is needed.
Where the displacement is experimentally measured, the measurement noise in the acceleration, which is amplified by differentiation with a small time-step, can be significantly reduced using physics-informed neural networks~\cite{davey2023measuring}.

The drag effects of superhydrophobic surfaces on spheres have been measured with this method using spheres with and without superhydrophobic surface treatment to determine the change in the terminal drag coefficient~\cite{mchale2009terminal,ahmmed2016internal,castagna2017super}.
The terminal Reynolds number can be changed by using spheres of different diameters and densities~\cite{mordant2000velocity,ahmmed2016internal}, which allows the Reynolds number effects on drag to be investigated.
Observations of the plastron formed over the sphere during these experiments show significant changes to its distribution over the surface~\cite{castagna2018wake}.
The plastron is deformed by the flow around the sphere such that it is more concentrated towards the rear of the sphere.
This results in a feedback loop between the flow in the wake of the sphere and its plastron, with the plastron distribution around the back of the sphere changing with the instantaneous pressure fluctuations, which affect the flow around the sphere.
Thus, the drag effects of superhydrophobic surface treatment on bluff body flows are not only a result of the change in conditions at the surface, but also the dynamic effects that the plastron has on the flow.

\subsection{Dynamic Mode Decomposition}

While turbulence often appears random, many turbulent flows contain coherent structures.
These structures are not always evident in the instantaneous measurements of the flow; however, they play a vital role in processes such as heat transfer and mixing, as well as affecting the dynamic behaviour of aerodynamic and hydrodynamic forces, such as lift and drag.
The identification of coherent structures and their effect on the flow can be used to create more parsimonious models of complex fluid dynamics problems by isolating those structures that have the greatest effect on the quantity of interest.
Thus, identifying coherent structures in turbulent flows is pertinent to many practical applications of fluid dynamics.

Modal decomposition is commonly used to determine the time information in patterns in turbulent fluctuations, which may indicate coherent structures and their timescale in the flow.
The basis of modal decomposition is the \ac{SVD}, which is also the basis of \ac{POD}~\cite{lumley1967structure,sirovich1987turbulence}.
The \ac{SVD} of the $m \times n$ matrix $\bs{X}$, where $\bs{X}_{m,n}$ represents a scalar field at time $t_m$ and spatial coordinate $\bs{x}_n$, is given by
\begin{equation}\label{eq:SVD}
  \bs{X} = \bs{U} \Sigma \bs{V}^T,
\end{equation}
where $\bs{U}$ and $\bs{V}^T$ contain the left and right eigenvectors of $\bs{X}$, respectively, and $\Sigma$ contains the corresponding singular values. 
For applications to turbulent flows, $\bs{X}$ is populated with the turbulent fluctuations, with the square of each singular value representing the turbulent kinetic energy of the corresponding row of $\bs{V}^T$.
Each row of $\bs{V}^T$ is an orthonormal spatial mode of $\bs{X}$, with temporal coefficients given by the corresponding column of $\bs{U}$.
The modes are sorted from largest to smallest $\Sigma$ and are thus ranked from the modes that contribute the most turbulent kinetic energy to $\bs{X}$ to those that contribute the least.

\ac{DMD} was first introduced by Schmid~\cite{schmid2010dynamic} to extract spatio-temporal structures from time-resolved data.
The matrix $\bs{A}$, such that
\begin{equation}\label{eq:DMDA}
  \bs{X}_{t + \Delta t} = \bs{A} \bs{X}_{t},
\end{equation}
where $\bs{X}_t$ is a vector field at time $t$ and $\bs{X}_{t + \Delta t}$ is the vector field at time $t + \Delta t$, and the matrix $\bs{A}$ describes the dynamical processes relating the vector field to the subsequent vector field, which is an approximation of the Koopman operator~\cite{tu2013dynamic}.
While originally formulated for time-resolved data, \ac{DMD} can be generalised to relate a set of snapshots pairs at time $t$ and $t+\Delta t$~\cite{tu2013dynamic}.
Unlike \ac{POD} modes, which represent correlations in space only, the structures of the \ac{DMD} modes are associated with specific oscillation frequencies and decay rates, and are thus directly linked to the system dynamics.
This allows the identification of the dominant frequencies, and the corresponding spatial structures~\cite{schmid2011applications}.

Unlike \ac{POD}, \ac{DMD} does not provide any ranking for modes, and thus makes isolation of the most important structures in the flow difficult without a priori knowledge about the dynamics of the flow.
In order to limit analysis to the most significant modes, Jovanovic et al.~\cite{jovanovic2014sparsity} formulated mode selection as a sparse optimisation problem, in what is known as \ac{SP}-\ac{DMD}.
\ac{SP}-\ac{DMD} simultaneously minimises the error between the \ac{DMD} reconstruction and the original data and the number of modes used in the reconstruction.
Thus, a minimal subset of dynamic modes, suitable for reduced-order modelling can be identified while preserving reconstruction accuracy.

Williams et al.~\cite{williams2015data} introduced \ac{EDMD} as a generalisation of standard \ac{DMD} in which the underlying dynamics are represented using a user-defined dictionary of nonlinear observables.
This formulation enables a richer approximation of nonlinear system behaviour and extends the range of dynamical features that can be extracted from snapshot data.
However, the computational cost of \ac{EDMD} increases rapidly with the size of the dictionary, particularly for high-dimensional datasets encountered in fluid mechanics.
To address this limitation, the authors subsequently proposed a kernel-based approach that implicitly defines the observable space through a kernel function, avoiding the explicit construction of large dictionaries while retaining the ability to approximate Koopman spectral quantities~\cite{williams2015kernel}.
Together, these methods improved the capability of \ac{DMD}-based techniques to analyse nonlinear, high-dimensional flow datasets and influenced many later developments in reduced-order modelling and data-driven flow analysis.

Beyond extending the observable space used to represent nonlinear dynamics, several developments focused on broadening the applicability and improving the robustness of \ac{DMD}.
Proctor et al.~\cite{proctor2016dynamic} introduced \ac{DMDc}, which incorporates actuation inputs into the decomposition process and enables the construction of data-driven reduced-order input-output models.
By separating the effects of forcing from the intrinsic flow dynamics, \ac{DMDc} extended the use of \ac{DMD} from flow analysis to flow-control and system-identification applications.
A comprehensive overview of these and other emerging developments was subsequently provided by Kutz et al.~\cite{kutz2016dynamic}, who unified connections between \ac{DMD}, Koopman operator theory, reduced-order modelling and machine learning, while surveying a growing range of \ac{DMD} variants and applications.

As the use of \ac{DMD} expanded to increasingly noisy experimental and numerical datasets, attention shifted towards improving the accuracy of modal estimates. Hemati et al.~\cite{hemati2017biasing} demonstrated that standard \ac{DMD} exhibits a systematic bias in the presence of measurement noise due to the asymmetric treatment of snapshot pairs.
To address this issue, they proposed \ac{TDMD}, a total least-squares formulation that accounts for errors in both past and future snapshots and yields more reliable estimates of modal characteristics from noisy data. Building on these efforts, Askham \& Kutz~\cite{askham2018variable} introduced \ac{OptDMD}, which employs a variable projection framework to fit all available snapshots simultaneously. Compared with standard formulations, \ac{OptDMD} reduces sensitivity to noise, accommodates unevenly sampled data and improves the estimation of modal frequencies and growth rates, making it a popular choice for data-driven analysis of complex fluid flows. Collectively, these developments have expanded the scope of \ac{DMD} from a modal decomposition technique to a versatile framework for modelling, control and spectral analysis, facilitating its application across a broad range of fluid-mechanics problems.

Beyond methodological developments, \ac{DMD} has been widely employed as a tool for identifying dynamically relevant coherent structures in complex flows.
Sayadi et al.~\cite{sayadi2014reduced} applied \ac{DMD} to transitional boundary-layer data to investigate near-wall structures during the late stages of transition, demonstrating its ability to isolate coherent features according to their characteristic frequencies rather than their energy content and thereby providing insight into the underlying transition mechanisms.
Chaugule et al.~\cite{chaugule2023investigating} combined dual-\ac{PIV} measurements with \ac{DMD} to investigate the near-field dynamics of a turbulent high-speed jet, extracting dominant coherent structures and their associated frequencies from high-spatial-resolution experimental data.
By comparing the spatial amplification of the identified modes with predictions from linear stability theory, they demonstrated that \ac{DMD} can successfully recover dynamically relevant flow structures embedded within turbulent jet flows.
Manganelli et al.~\cite{manganelli2025effect} employed sparsity-promoting \ac{DMD} to examine the influence of Coriolis-induced wind veer on the wake dynamics of a utility-scale wind turbine.
Their analysis identified the dominant coherent structures governing wake recovery and enabled the construction of a reduced-order representation of the wake using a limited subset of dynamically significant modes.
These studies illustrate the versatility of \ac{DMD} as a flow-analysis tool, spanning applications from canonical transitional boundary layers and turbulent jets to complex atmospheric and wind-energy flows.

\subsection{The Current Work}
The current work investigates the effect of superhydrophobic surface treatment on the dynamics of the shear layer in the near wake of a sphere, where the plastron is actively sustained by supplying air through pores in the surface of the sphere.
Three spheres were manufactured: (i) a smooth untreated sphere to serve as a reference, (ii) an untreated sphere with a porous shell to determine the change in dynamics from the addition of the pores, and (iii) a porous sphere with superhydrophobic surface treatment to determine the change in the flow dynamics when the plastron is sustained.
The instantaneous velocities in a plane aligned with the centre of each sphere were measured using \ac{PIV}, from which the instantaneous velocity fluctuations were determined.
Spatial \ac{DMD} was performed in the streamwise direction on the velocity fluctuations in the shear layer of each sphere.
The resulting \ac{DMD} modes were sorted in order of descending energy share.
The \ac{DMD} modes for each sphere were compared to determine the effect of the superhydrophobic surface treatment on the dynamics when the plastron is sustained.

\section{Methodology}
\subsection{Experimental Facility}
\subsubsection{Vertical Water Tunnel}

\begin{figure*}[htbp]
  \centering
  \includegraphics[width=6.4in]{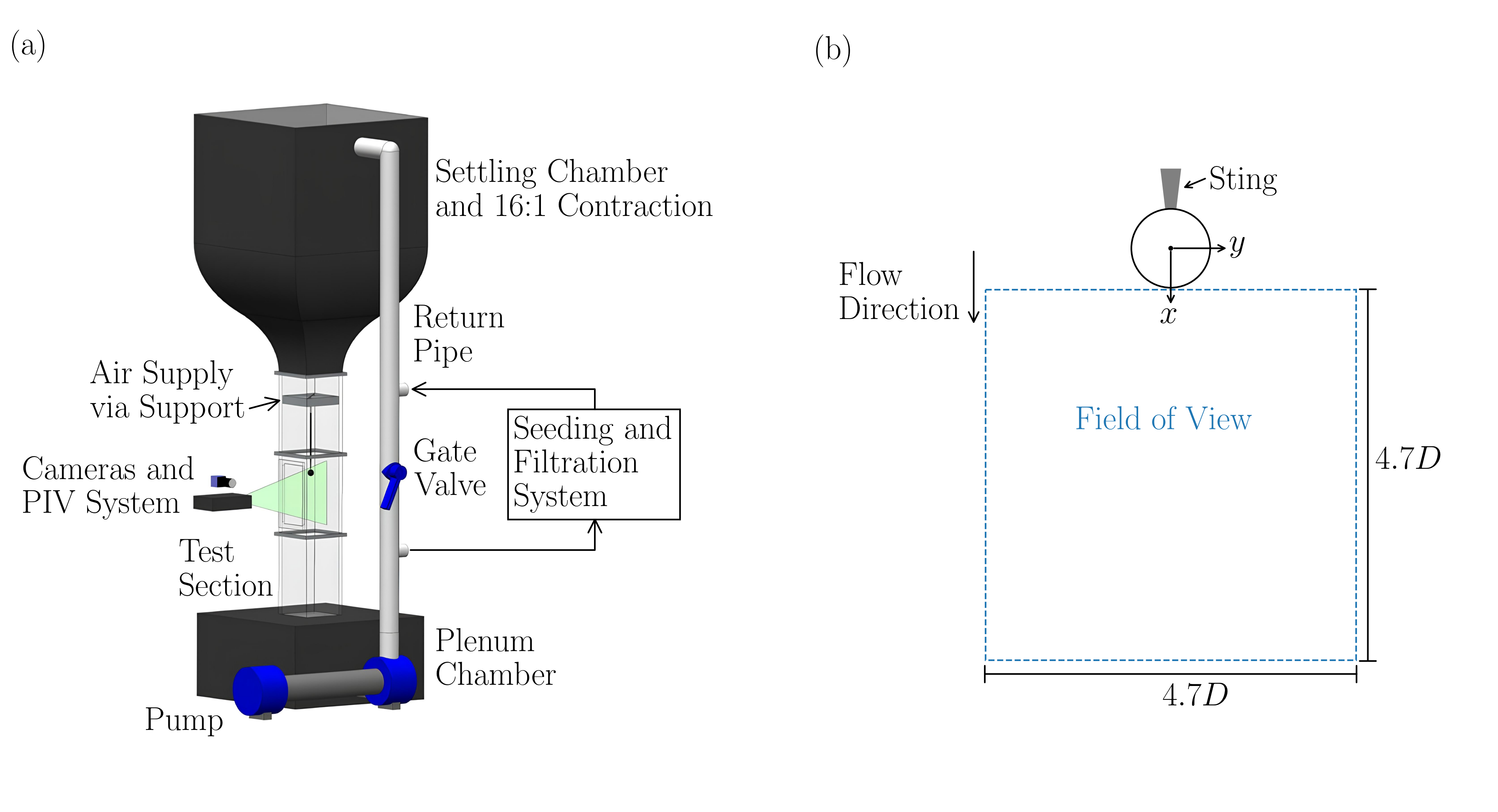}
  \caption{(a) Vertical water tunnel facility and (b) experimental field of view (reproduced from~\cite{davey2025experimental}).}\label{fig:VWT}
\end{figure*}

The flow in the near wake of the spheres was measured in a vertical water tunnel facility with a 250 mm $\times$ 250 mm cross-section and a 1.5 m working section~\cite{gordon2004investigation,buchner2015measurements,davey2025experimental}, as shown in figure~\ref{fig:VWT}a.
The settling chamber above the working section includes a 16:1 contraction and honeycomb screens to reduce the turbulence in the working section.
The 0.5 m high test section is located in the middle of the working section, with a cut-out section on one wall to provide access.
An acrylic panel is used to seal the tunnel, and is flush with the internal side of the wall to minimise the disturbance to the flow.
Water is pumped from the plenum chamber below the working section up to the settling chamber through a return pipe, such that the flow through the working section is downwards.
Seeding of the flow and the filtration of the tunnel are performed using an auxiliary circuit connected to the return pipe, which is isolated from the tunnel during experiments.

\begin{figure}[htbp]
  \centering
  \includegraphics[width=3in]{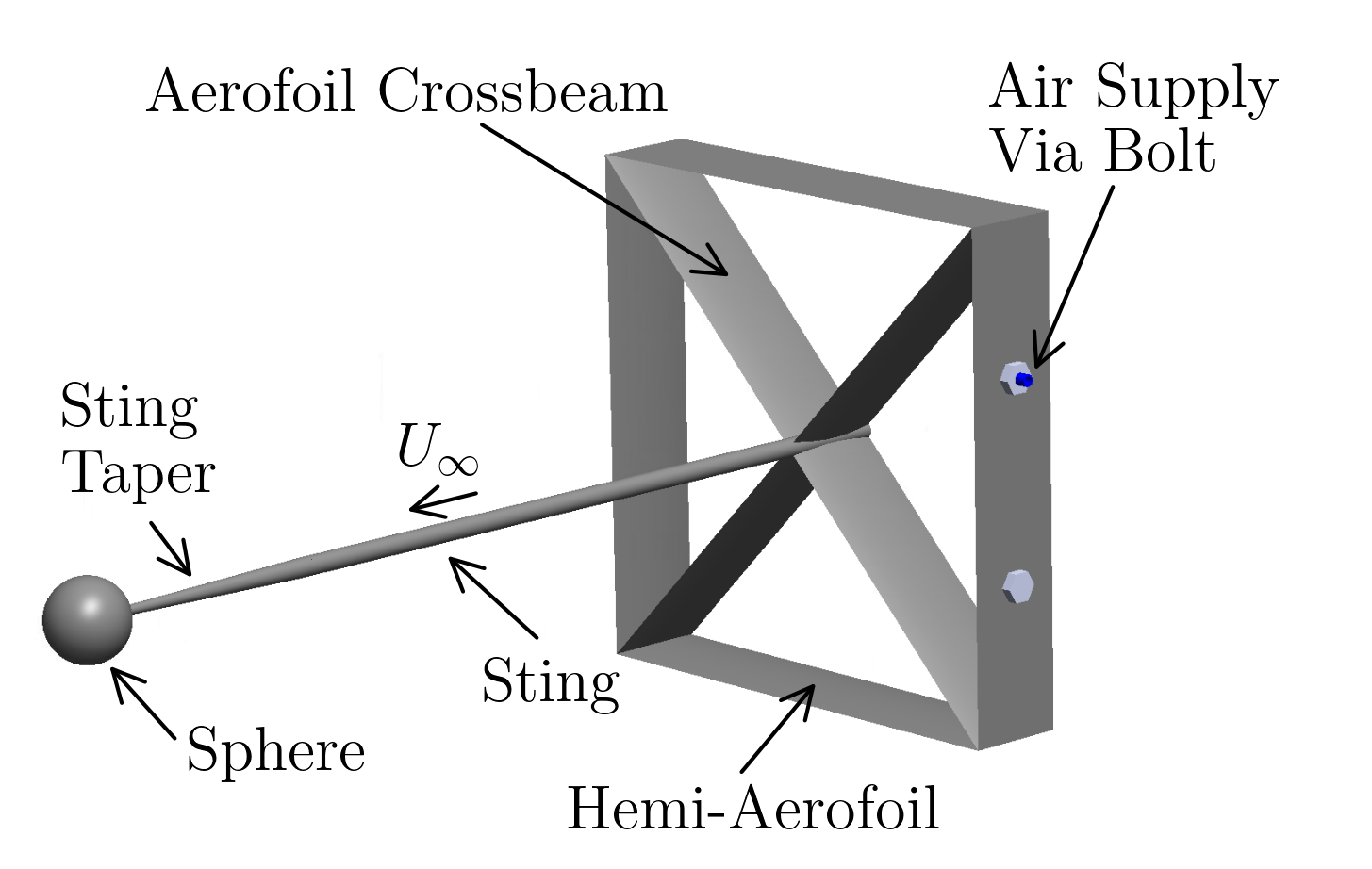}
  \caption{Aerofoil crossbeam and sting used for mounting the sphere (reproduced from~\cite{davey2025experimental}).}\label{fig:Cross}
\end{figure}

A crossbeam structure of thin aerofoils is mounted above the test section to support the sting to which the sphere is attached.
The sting has a diameter of 9.3 mm until the last 100 mm, over which it tapers linearly to 4.65 mm and ends in a threaded rod for mounting the sphere.
The sting and the supporting aerofoil structure include an internal channel connected to a compressed air supply, through which air was supplied to the \ac{SHS} sphere to maintain the plastron.
A close-up view of the crossbeam structure, sting and sphere is shown in figure~\ref{fig:Cross}.

\subsubsection{Sphere Design}

The spheres were 3-D printed with a resolution of 22 \textmu\@m $\times$ 22 \textmu\@m in the horizontal plane and 35 \textmu\@m in the vertical direction.
The spheres were printed with a diameter of $D=40$ mm and a 2.5 mm-thick shell.
An internal support structure was included to support the sphere during printing and the experiments, with slits to allow air to flow from the sting to the surface of the sphere.
An external boss was included on the top of the sphere to smooth the transition from the sting to the sphere.

\begin{figure}[htbp]
  \centering
  \includegraphics[width=6.4in]{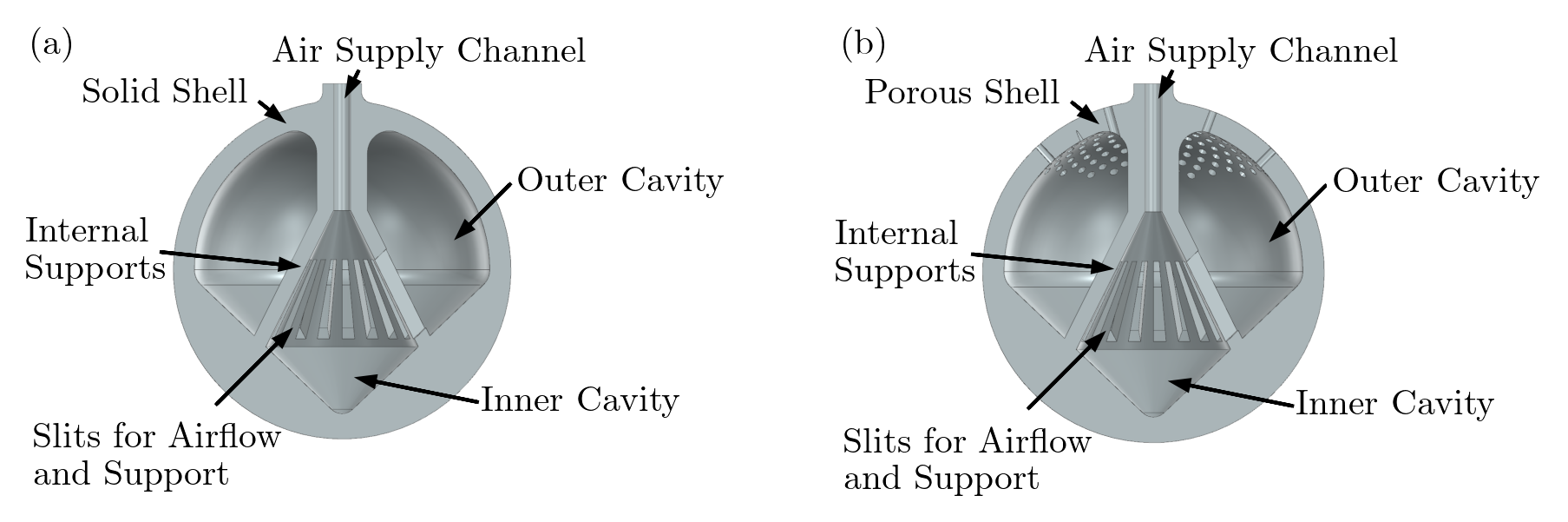}
  \caption{Cut-away view of 3D-printed sphere models for the (a) smooth reference sphere and (c) porous spheres with and without superhydrophobic surface treatment (reproduced from~\cite{davey2025experimental}).}\label{fig:spheres}
\end{figure}

In order to allow air to reach the surface of the sphere, pores with a diameter of 1 mm were placed on the sphere's surface according to the angular position given by
\begin{equation}\label{eq:Pores}
  \begin{bmatrix}
    \theta \\ \phi	
  \end{bmatrix}_i
  =
  \begin{bmatrix}
    \frac{2 \pi i}{\Phi_{GR}} \\ \\ \cos^{-1} \left(\frac{1 - 2(i+\frac{1}{2})}{N_P} \right)
  \end{bmatrix},		
\end{equation}
where $\theta$ and $\phi$ are the azimuthal and polar angles, respectively, $N_P$ is the total number of pores over the sphere, and $\Phi_{GR} = \frac{1 +\sqrt{5}}{2}$ is the golden ratio~\cite{davey2025experimental}.
The number of points was chosen to achieve a porosity of 20\%.
The pores that would interfere with the external boss were removed, as were those positioned further than 50$^\circ$ from the top of the sphere.
The latter condition was chosen to prevent escaping bubbles from reflecting laser light into the camera.
A sphere with no pores or surface treatment was printed to serve as a \ac{REF}, and a \ac{PRS} sphere was printed and left without surface treatment to determine the effect of the pores on the flow in the absence of superhydrophobic surface treatment.
A second porous sphere was treated with Rust-Oleum NeverWet\textsuperscript{TM} \ac{SHS} coating to determine the effect of the sustained plastron on the flow.
The spheres are shown in figure~\ref{fig:spheres}.

\subsection{Experimental Method}

The freestream velocity of the tunnel was set to $U_\infty=200$ mm/s for all spheres, resulting in a Reynolds number of $Re_D=7,780$, with a corresponding background turbulence of 2\%~\cite{davey2025experimental}.
The mean convective timescale arising from the sphere diameter and freestream velocity is $t_c=0.2$ s~\cite{davey2025experimental}. 
The flow was seeded with glass spheres of 11\textmu\@m diameter, and the single-exposed particle images were illuminated by a pair of 400 mJ Spectra Physics Nd:YAG lasers firing 7 ns pulses and recorded using a PCO Panda 26 DS camera.
The camera consists of a 5120 $\times$ 5120 array of 2.5 \textmu\@m $\times$ 2.5 \textmu\@m pixels, which was used to capture a 190 mm $\times$ 190 mm field of view beginning directly below the sphere and aligned to its centre, as shown in figure~\ref{fig:VWT}b, with a resulting spatial resolution of 37 \textmu\@m/px.
Pairs of single-exposed images were captured with a separation of $\Delta t = 1$ ms, and image pairs were acquired at a rate of 1 Hz, resulting in a separation between instantaneous velocity fields of $5t_c$.
The acquisition of $N_T > 10,000$ images for each sphere was synchronised with the laser pulses using a \ac{BBB} pulse generator~\cite{fedrizzi2015application}.

\begin{figure}
  \centering
  \includegraphics[width=2in]{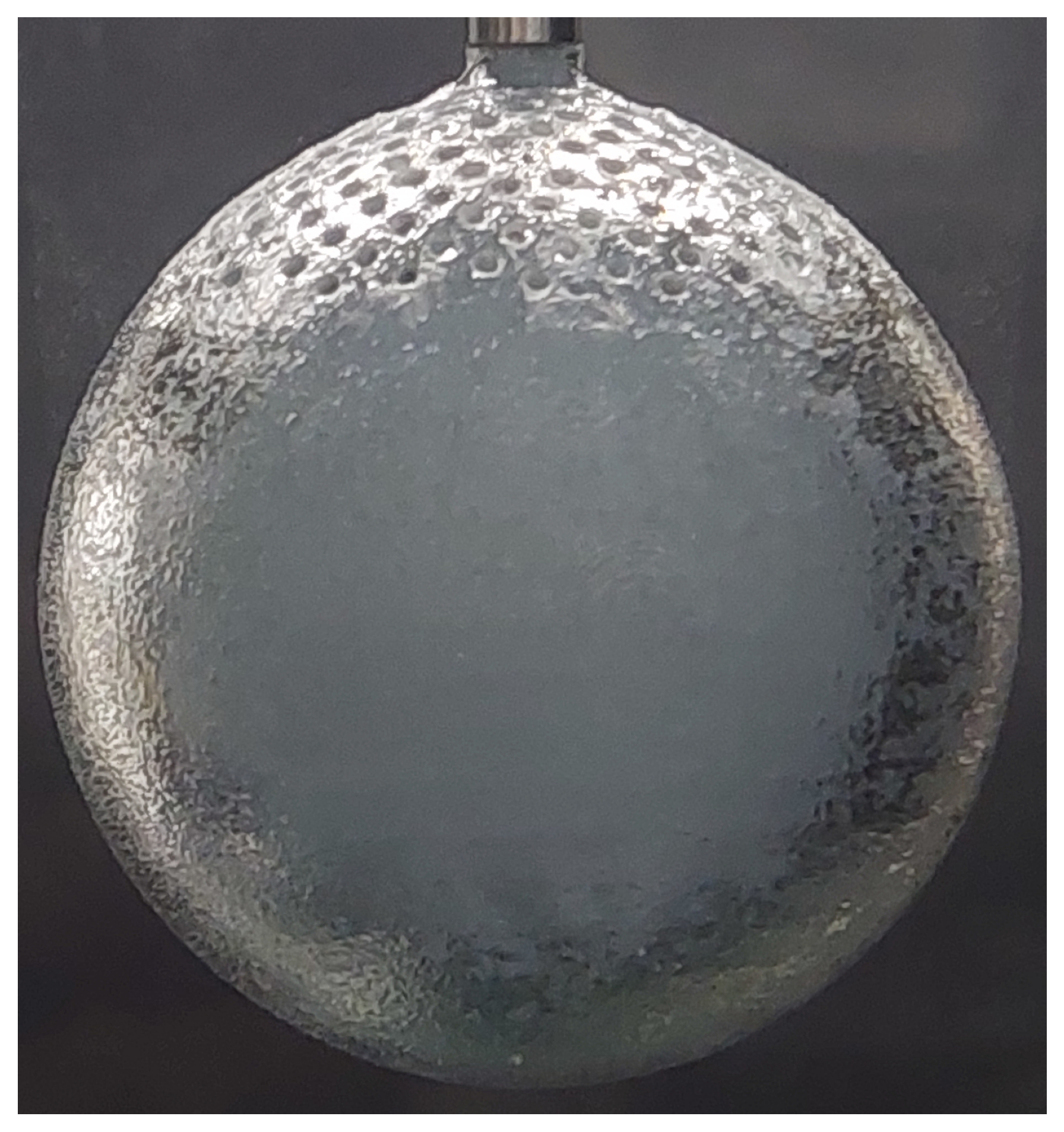}
  \caption{Image of the plastron formed on the superhydrophobic sphere, taken with a non-\ac{PIV} camera (reproduced from~\cite{davey2025experimental}).}\label{fig:plastron}
\end{figure}

In order to sustain the plastron formed on the \ac{SHS} sphere for the duration of the experiments, as it would otherwise be depleted over time by the forces generated on it by the flow, air was supplied to the surface of the sphere via pores in the sphere's surface~\cite{davey2025experimental}.
The air was supplied at very low pressure, with a flow rate of 0.18$\pm$0.006 litres per minute to prevent air bubbles from breaking off at the surface and disturbing the oncoming flow.
To prevent blockage of the pores, settled seeding particles were cleared from the pores periodically.
No roughness was added to the sphere, and as a result the plastron was thinner than those in settling sphere experiments with added roughness~\cite{castagna2018wake}, as shown in figure~\ref{fig:plastron}.
This reduced the risk of reflections into the camera in the field of view.

Measurement of the instantaneous streamwise and transverse velocities, $u(x,y,t)$ and $v(x,y,t)$, respectively, in the wake of each sphere was performed using \ac{2C-2D} \ac{MCCD}-\ac{PIV}~\cite{davey2025experimental,soria1996investigation}.
In order to balance the intensity of the single-exposed images, the mean background images of the first and second exposures were subtracted from the corresponding images before processing.
The final interrogation window was 64 px $\times$ 32 px, with a grid spacing of 32 px $\times$ 16 px in $x$ and $y$, respectively.
The instantaneous velocity fluctuations were determined for each vector field by subtracting the mean velocity field from the instantaneous velocity fields.
The key experimental parameters are summarised in table~\ref{tab:ExperimentalParameters}.

\begin{table*}[htbp]
  \centering
  \caption{Table of experimental parameters~\cite{davey2025experimental}.}\label{tab:ExperimentalParameters}
  \begin{tabular}{l c c c}
    \toprule
    Parameter                       & Symbol                             & Physical Units         & Relative Units            \\
    \midrule
    Freestream velocity             & $U_\infty$                         & 200 mm/s               & $-$                       \\
    Sphere diameter                 & $D$                                & 40 mm                  & $-$                       \\
    Reynolds number                 & $Re_D$                             & 7,780                  & $-$                       \\
    Time between laser pulses       & $\Delta t$                         & 1 ms                   & $0.005t_c$                \\
    Time between velocity fields    & $\Delta T$                         & 1 s                    & $5t_c$                    \\
    Spatial resolution              & $SR$                               & 37 \textmu\@m/px       & 9.25$\times10^{-4} D$     \\
    Interrogation window            & $-$                                & 64 px $\times$ 32 px   & 0.06$D$ $\times$ 0.03$D$  \\
    Grid spacing                    & $-$                                & 32 px $\times$ 16 px   & 0.03$D$ $\times$ 0.015$D$ \\
    Field of view                   & $-$                                & 190 mm $\times$ 190 mm & 4.7$D$ $\times$ 4.7$D$    \\
    Number of 2C-2D velocity fields & $N_T$                              & $>$10,000              &                           \\
    \bottomrule
  \end{tabular}
\end{table*}

The uncertainty of the \ac{2C-2D}-\ac{MCCD}\ac{PIV} algorithm used is 0.06 px~\cite{soria1996investigation} at the 95\% confidence level, which corresponds to a single-sample uncertainty of 2.2 mm/s in the instantaneous velocity $u$.
The mean of 10,000 instantaneous velocity vectors, $\overline{u}$, therefore has an uncertainty of 0.04 mm/s, accounting for the streamwise Reynolds stress~\cite{sun20252c2d}.
As the instantaneous velocity fluctuations are given by
\begin{equation}\label{eq:uf}
  u' = u - \overline{u},
\end{equation}
their uncertainty $\sigma_{u'}$ is given by
\begin{equation}\label{eq:sigma_uf}
  \sigma_{u'}^2 = \sigma_{u}^2 + \sigma_{\overline{u}}^2,
\end{equation}
and is equal to 1.2\% of the freestream velocity at the 95\% confidence level. 

\subsection{Spatial Dynamic Mode Decomposition}
In order for \ac{DMD} to be meaningful, the $\Delta t$ between vector fields must be adequately small to capture the dynamics of the flow from which the vector fields are taken, which limits its application to time-resolved~\cite{schmid2011applications} or pairwise time-resolved data~\cite{chaugule2023investigating}.
When the vector fields are not separated by sufficiently small $\Delta t$ for \ac{DMD} to be performed in time, the matrices $\bs{X}_{t}$ and $\bs{X}_{t + \Delta t}$ can be replaced with $\bs{X}_{x}$ and $\bs{X}_{x + \Delta x}$, respectively, where $\Delta x$ is the separation in the streamwise direction~\cite{schmid2010dynamic}.
Thus, the dynamical processes being represented by the \ac{DMD} are
\begin{equation}\label{eq:sDMD}
  \bs{X}_{x + \Delta x} = \bs{A} \bs{X}_{x},
\end{equation}
where $\bs{A}$ represents the dynamical processes between the streamwise positions $x$ and $x + \Delta x$.

Each row of $\bs{X}_x$ represents an independent realisation, or snapshot, of the turbulent fluctuations in the flow.
The corresponding row of $\bs{X}_{x+\Delta x}$ represents the same realisation with a streamwise shift of $\Delta x$.
This serves as an analogy for the dual-PIV application by Chaugule et al.~\cite{chaugule2023investigating}, with the streamwise shift taking the place of the temporal separation used in that work.
Thus, this approach does not provide a model for the temporal evolution of the flow.
Instead these modes represent the spatial evolution of the flow, with the eigenvalues representing the spatial decay and wavelength of the modes.

The matrix $\bs{X}_x$ is defined using the instantaneous velocity fluctuations at all streamwise locations except for the furthest downstream, where each row represents an instant in time and each column represents a position in space.
In order for both velocity components to be included simultaneously, the matrix $\bs{X}_x$ is defined as
\begin{equation}\label{eq:Xx}
  \bs{X}_{x} = \begin{bmatrix}
    u'\big({(x,y)}_1,t_1\big) & \cdots & u'\big({(x,y)}_N,t_1\big) & v'\big({(x,y)}_1,t_1\big) & \cdots & v'\big({(x,y)}_N,t_1\big)\\
    \vdots & \ddots & \vdots & \vdots & \ddots & \vdots & \\
    u'\big({(x,y)}_1,t_{N_T}\big) & \cdots & u'\big({(x,y)}_N,t_{N_T}\big) & v'\big({(x,y)}_1,t_{N_T}\big) & \cdots & v'\big({(x,y)}_N,t_{N_T}\big)\\
  \end{bmatrix},
\end{equation}
where $t_m$ is the time of the $m^{th}$ velocity field, ${(x,y)}_n$ is the $n^{th}$ spatial coordinate, and $N$ denotes the total number of spatial coordinates excluding those at the furthest downstream location.
The matrix $\bs{X}_{x + \Delta x}$ is thus defined by replacing $x$ with $x + \Delta x$ in equation (\ref{eq:Xx}).
Direct calculation and manipulation of $\bs{A}$ in (\ref{eq:DMDA}) is computationally cumbersome for large data sets.
Thus, exact \ac{DMD} uses the reduced \ac{SVD} of $\bs{X}_{t}$ to define the matrix
\begin{equation}\label{eq:A_tilde}
  \bs{\wt{A}} = \bs{U}^T_r \bs{X}_{t + \Delta t} \bs{V}_r \Sigma^{-1}_r,
\end{equation}
where $r$ denotes the rank of the reduced-order \ac{SVD}~\cite{tu2013dynamic}.
This is used to compute the \ac{DMD} modes and their eigenvalues without requiring direct computation or manipulation of $\bs{A}$.

The reduced \ac{SVD} of $\bs{X}_x$ is given by
\begin{equation}\label{eq:rSVD}
  {(\bs{X}_x)}_r = \bs{U}_r \Sigma_r \bs{V}^T_r,
\end{equation}
where $r$ is the rank of the reduction.
The matrix $\bs{\wt{A}}$ is computed using equation (\ref{eq:A_tilde}), with $\bs{X}_{x}$ and $\bs{X}_{x + \Delta x}$ used in place of $\bs{X}_{t}$ and $\bs{X}_{t + \Delta t}$, respectively.

The eigendecomposition of $\bs{\wt{A}}$
\begin{equation}\label{eq:EigenDecompAtilde}
  \bs{\wt{A}}\bs{\Phi} = \wt{\mu} \bs{\Phi},
\end{equation}
yields eigenvalues $\wt{\mu}$ and eigenvectors $\bs{\Phi}$, from which the $k^{th}$ \ac{DMD} mode of $\bs{\wt{A}}$ is given by
\begin{equation}\label{eq:DMDmodes}
  \psi_k(x,y) = \wt{\mu}_k {\bs{\Phi}}^{-1}_k \Sigma_r \bs{V}_r^T.
\end{equation}
The eigenvalues of $\bs{A}$, $\mu$, are related to those of $\bs{\wt{A}}$ by
\begin{equation}\label{eq:eigenvaluesA}
  \mu_k = \frac{\log(\wt{\mu}_k)}{\Delta x}.
\end{equation}
The amplitudes of each \ac{DMD} mode for each realisation are given by~\cite{chaugule2023investigating}
\begin{equation}\label{eq:DMDamps}
  b_k(t) = \bs{X}_{x+\Delta x} \bs{V}_r^T \bs{\Sigma}_r^{-1} \bs{\Phi}_r \wt{\mu}_k^{-1}.
\end{equation}
The scale-invariant energy contribution of each mode is given by
\begin{equation}\label{eq:DMDenergy}
  E_k = ||\psi_k||_2^2 ||b_k||_2^2,
\end{equation}
where
\begin{equation}\label{eq:Euclidean}
  ||\ldots||_2 =   \sqrt{ {(\ldots)}^* (\ldots)}
\end{equation}
is the Euclidean norm, with $^*$ denoting the conjugate transpose.
The energy share of each mode is
\begin{equation}\label{eq:EnergyShare}
  s_k = \frac{E_k}{\sum_{k=1}^r E_k}.
\end{equation}

The growth/decay rate of the $k^{th}$ \ac{DMD} mode in the streamwise direction, $r^x_k$, is given by the real part of $\mu_k$
\begin{equation}\label{eq:lambdaReal}
  r^x_k = \Re(\mu_k) = \frac{\log(|\wt{\mu}_k|)}{\Delta x},
\end{equation}
and the wavenumber is given by the imaginary part
\begin{equation}\label{eq:lambdaImag}
  \Im(\mu_k) = \frac{\theta_{\wt{\mu}_k}}{\Delta x},
\end{equation}
where $|\wt{\mu}_k|$ and $\theta_{\wt{\mu}_k}$ are the magnitude and angle of $\wt{\mu}_k$ in the complex plane, respectively.
The wavenumber defined in equation (\ref{eq:lambdaImag}) is in radians per metre and is divided by $2\pi$ to be converted to m$^{-1}$.
Therefore, the wavelength of the $k^{th}$ \ac{DMD} mode in the streamwise direction, $\lambda^x_k$ is
\begin{equation}\label{eq:wavelength}
  \lambda_k^x = \frac{2\pi}{\Im(\mu_k)} = \frac{2 \pi\Delta x}{\theta_{\wt{\mu}_k}}.
\end{equation}

\subsection{Application of Dynamic Mode Decomposition to the Flow Over Spheres}
In order to capture the dynamics of the shear layer on either side of the sphere, the \ac{DMD} was performed in the domains $y/D \in [0,1]$ and $y/D \in [-1,0]$ independently.
This centres each domain on the corresponding side of the sphere, and maintaining the independence of the analysis of each side of the sphere means that the assumption of axisymmetry is not required.
This is particularly important in the case of the \ac{PRS} and \ac{SHS} spheres, where the addition of the pores and superhydrophobic surface treatment may result in asymmetries.

\begin{figure*}[htbp]
  \centering
  \includegraphics[width=6.4in]{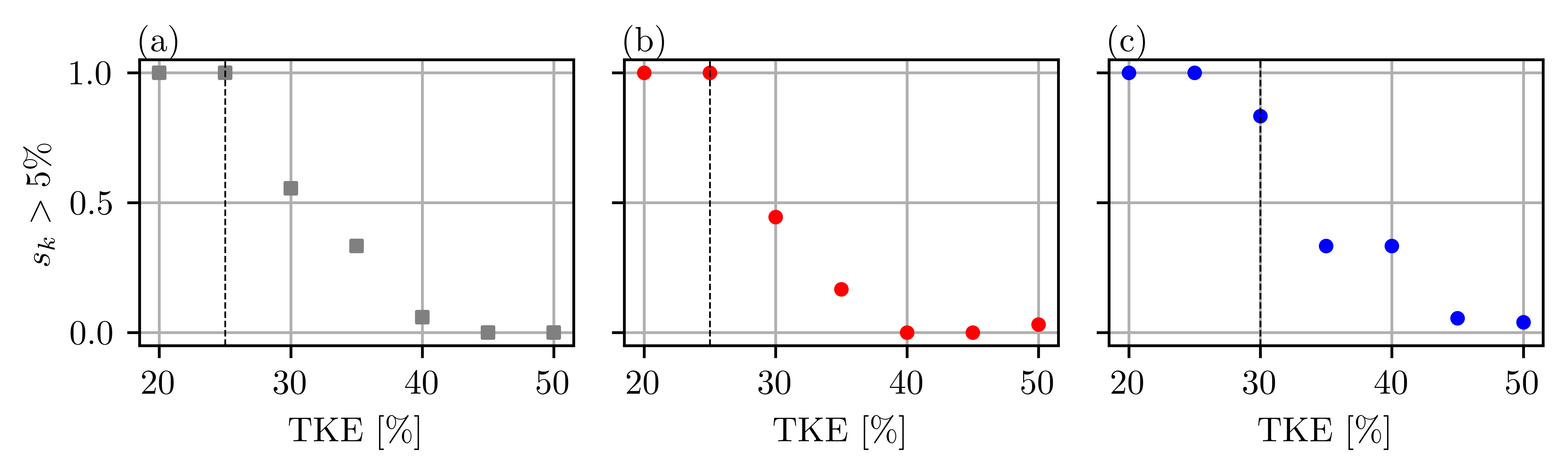}
  \caption{Proportion of \ac{DMD} modes with an energy share of 5\% or more for each cumulative \ac{TKE} for the (a) REF, (b) PRS and (c) SHS spheres.
           Black dotted lines indicate the chosen \ac{TKE} for each sphere.}\label{fig:DMDenergy}
\end{figure*}

As the flow in the wake of the sphere is highly three-dimensional, rank reduction of the \ac{SVD} was necessary to limit the \ac{DMD} to the most dominant modes in the measurement plane and avoid highly three-dimensional and small-scale modes that are not well-resolved by the measurement plane.
Thus, the rank of the \ac{SVD} was reduced to the \ac{POD} modes which contributed a cumulative \ac{TKE} of 25\% to 50\% in 5\% increments.
The energy share of the \ac{DMD} modes at each rank was determined from their amplitudes using equations (\ref{eq:DMDenergy}) to (\ref{eq:EnergyShare}).
The proportion of \ac{DMD} modes in both domains with an energy share of greater than 5\% for each case is shown for each \ac{TKE} in figure~\ref{fig:DMDenergy}.

The proportion of modes with an energy share of 5\% drops off sharply for a cumulative \ac{TKE} of 25\% for the \ac{REF} and \ac{PRS} spheres and 30\% for the \ac{SHS} sphere, shown in figure~\ref{fig:DMDenergy} for each sphere.
These levels of cumulative \ac{TKE} are achieved by including 12 \ac{SVD} modes for each sphere.
Thus, a rank of 12 was used for the reduced \ac{SVD} used in the \ac{DMD} for all three spheres.
The resulting \acp{RRR} were 0.21, 0.20 and 0.15 for the \ac{REF}, \ac{PRS} and \ac{SHS} spheres, respectively.
These parameters are summarised in table~\ref{tab:DMDparameters}.

\begin{table}[htbp]
  \centering
  \caption{Table of \ac{DMD} parameters for each sphere.}\label{tab:DMDparameters}
  \begin{tabular}{l c c c}
    \toprule
    Dataset	& TKE Ratio	& Rank	& RRR  \\
    \midrule
    REF     & 0.25      & 12    & 0.21 \\
    PRS     & 0.25      & 12    & 0.20 \\
    SHS     & 0.30      & 12    & 0.15 \\
    \bottomrule
  \end{tabular}
\end{table}

The \ac{DMD} results for the domains above and below the centreline were related to one another by matching eigenvalues using the Hungarian method~\cite{kuhn1955hungarian}.
Once the corresponding modes in each domain were identified, the cross-domain complex correlation coefficient
\begin{equation}\label{eq:CompCorr+-}
  \mc{R}_{+,-}^{C} = \frac{\langle \psi^+, \psi^- \rangle}{||\psi^+||_2 ||\psi^-||_2}, 
\end{equation}
where $^+$ and $^-$ denote the mode shapes above and below the centreline, respectively, was calculated.
The angle $\angle\mathcal{R}_{+,-}^\mathbb{C}$ and magnitude $|\mathcal{R}_{+,-}^\mathbb{C}|$ of the cross-domain complex correlation coefficient gives the phase-offset and corresponding cross-correlation coefficient, respectively.
The \ac{DMD} mode shapes, eigenvalues and energy shares were then determined by the average of the corresponding modes in each plane.

The energy share was used to rank the modes for each sphere, and the resulting eigenvalues and energy shares are shown in figure~\ref{fig:Eigs}.
The out-of-plane vorticity
\begin{equation}\label{eq:vorticity}
\omega = \frac{\partial v}{\partial x} - \frac{\partial u}{\partial y},
\end{equation}
of each mode was calculated from its streamwise $u$ and transverse $v$ components to provide an additional visualisation of the structure of each mode.
In order to find comparable modes between the spheres, the complex correlation coefficient between the $i^{th}$ and $j^{th}$ \ac{DMD} modes of each sphere
\begin{equation}\label{eq:CompCorrij}
  \mc{R}_{i,j}^{C} = \frac{\langle \psi_i, \psi_j \rangle}{||\psi_i||_2 ||\psi_j||_2}, 
\end{equation}
was determined, where $\psi_i$ and $\psi_j$ include both the streamwise and transverse fluctuations of the corresponding modes.
Modes with a wavelength longer than twice the measurement domain were moved to the end of the mode ranking, as these modes may not represent dynamics resolvable within the current measurement domain.

\section{Results}

\begin{table*}[h]
  \centering
  \caption{Magnitudes $|\mathcal{R}_{+,-}^\mathbb{C}|$ and angles $\angle\mathcal{R}_{+,-}^\mathbb{C}$ of the cross-domain complex correlation coefficients of the \ac{DMD} modes of each sphere.}\label{tab:Corr}
  \begin{tabular}{l c r c r c r}
    \toprule
        & \multicolumn{2}{c}{\ac{REF}}
        & \multicolumn{2}{c}{\ac{PRS}}
        & \multicolumn{2}{c}{\ac{SHS}} \\
    \midrule
    $k$ & $|\mathcal{R}_{+,-}^\mathbb{C}|$ & $\angle\mathcal{R}_{+,-}^\mathbb{C}$
        & $|\mathcal{R}_{+,-}^\mathbb{C}|$ & $\angle\mathcal{R}_{+,-}^\mathbb{C}$
        & $|\mathcal{R}_{+,-}^\mathbb{C}|$ & $\angle\mathcal{R}_{+,-}^\mathbb{C}$ \\
    \midrule
    1   & 0.96 & -0.96$\pi$ & 0.89 & -0.95$\pi$ & 0.85 &  0.88$\pi$ \\
    2   & 0.98 &  0.03$\pi$ & 0.95 &  0.07$\pi$ & 0.77 &  0.39$\pi$ \\
    3   & 0.99 &  0.99$\pi$ & 0.96 &  0.00$\pi$ & 0.90 &  0.82$\pi$ \\
    4   & 0.98 &  0.48$\pi$ & 0.97 & -0.08$\pi$ & 0.96 &  0.94$\pi$ \\
    5   & 0.95 &  0.42$\pi$ & 0.80 &  0.73$\pi$ & 0.89 &  0.43$\pi$ \\
    6   & 0.96 & -0.72$\pi$ & 0.94 &  0.94$\pi$ & 0.88 &  0.57$\pi$ \\
    \bottomrule
  \end{tabular}
\end{table*}

The modes of the \ac{REF} sphere have consistently high $|\mathcal{R}_{+,-}^\mathbb{C}|$, as shown in table~\ref{tab:Corr}, with the lowest being 0.95 for the fifth mode, suggesting that these modes are highly symmetric/antisymmetric across the centreline.
The \ac{PRS} modes also have high $|\mathcal{R}_{+,-}^\mathbb{C}|$, although the values for the first and fifth modes are noticeably lower than the rest at 0.89 and 0.80, respectively.
The values of $|\mathcal{R}_{+,-}^\mathbb{C}|$ for the \ac{SHS} sphere are more varied, ranging from 0.77 for the second mode to 0.96 for the fourth mode.
This suggests that the symmetry/antisymmetry of these modes is affected by the addition of superhydrophobic surface treatment.

\begin{figure*}[h]
  \centering
  \includegraphics[width=6.4in]{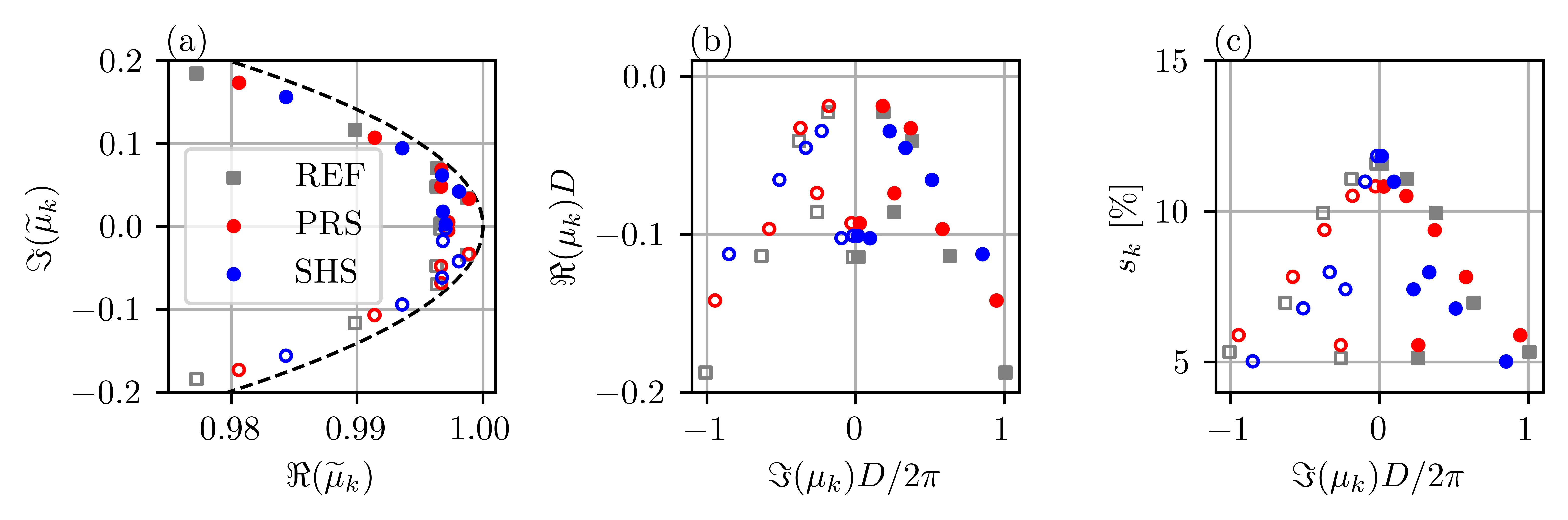}
  \caption{\ac{DMD} eigenvalues of (a) $\wt{\bs{A}}$, (b) $\bs{A}$, and (c) energy shares $s_k$ for each sphere.
           Hollow symbols denote conjugate pairs.
           The dashed line in (a) represents the unit circle.}\label{fig:Eigs}
\end{figure*}

The $\wt{\mu}_k$ for all three spheres sit inside the unit circle, as shown in figure~\ref{fig:Eigs}a, and thus the corresponding $\mu_k$ have negative real parts, as shown in figure~\ref{fig:Eigs}b.
This is consistent with the turbulence decaying downstream of the sphere.
The energy shares of the modes are similar between the \ac{REF} and \ac{PRS} spheres, suggesting that the effect of the pores on the dynamic modes is relatively minor compared to the effect of the superhydrophobic surface treatment, as shown in figure~\ref{fig:Eigs}c.

\begin{table*}[h]
  \centering
  \caption{Energy shares $s_k$, decay rates $r_k^x$ and wavelengths $\lambda_k^x$ of the \ac{DMD} modes of each sphere.}\label{tab:DMDEigs}
  \begin{tabular}{l c c c  c c c  c c c c}
    \toprule
        & \multicolumn{3}{c}{\ac{REF}} & \multicolumn{3}{c}{\ac{PRS}} & \multicolumn{3}{c}{\ac{SHS}} \\
    \midrule
    $k$ &  $s_k$ [\%] &  $r_k^x D$ &  $\lambda_k^x/D$
        &  $s_k$ [\%] &  $r_k^x D$ &  $\lambda_k^x/D$
        &  $s_k$ [\%] &  $r_k^x D$ &  $\lambda_k^x/D$ \\
    \midrule
    1   &  22.14 &  -0.02 &   5.38 & 21.03 &  -0.02 &	  5.54 & 15.96 &  -0.05 &   2.99 \\
    2   &  19.88 &  -0.04 &   2.64 & 18.76 &  -0.03 &	  2.70 & 14.82 &  -0.03 &   4.38 \\
    3   &  13.91 &  -0.11 &   1.58 & 15.65 &  -0.10 &	  1.72 & 13.56 &  -0.07 &   1.95 \\
    4   &  10.66 &  -0.19 &   0.99 & 11.78 &  -0.14 &	  1.06 & 10.03 &  -0.11 &   1.18 \\
    5   &  10.24 &  -0.09 &   3.86 & 11.12 &  -0.07 &	  3.84 & 23.67 &  -0.10 &  65.38 \\
    6   &  23.17 &  -0.11 &  53.89 & 21.65 &  -0.09 &	 36.33 & 21.96 &  -0.10 &  10.44 \\
    \bottomrule
  \end{tabular}
\end{table*}

Both the \ac{REF} and \ac{PRS} spheres feature five \ac{DMD} modes with wavelengths shorter than twice the measurement domain and one \ac{DMD} mode with a wavelength significantly longer than the measurement domain.
Each of the first five modes of these spheres exhibit similar energy shares, decay rates and wavelengths, as shown in table~\ref{tab:DMDEigs}.
Only four of the \ac{SHS} modes have wavelengths shorter than twice the measurement domain, with the fifth mode having a wavelength significantly above this threshold and the sixth mode being only slightly above, at 65.38$D$ and 10.44$D$, respectively.
The sixth \ac{REF} and \ac{PRS} modes each have the largest energy share, despite the wavelength of these modes being significantly longer than the measurement domain, at 53.89$D$ and 36.33$D$, respectively.
This is also the case for the fifth and sixth \ac{SHS} modes.
Despite all spheres exhibiting modes with long wavelengths, the values of these wavelengths vary significantly between spheres.

\begin{figure*}[h]
  \centering
  \includegraphics[width=6.4in]{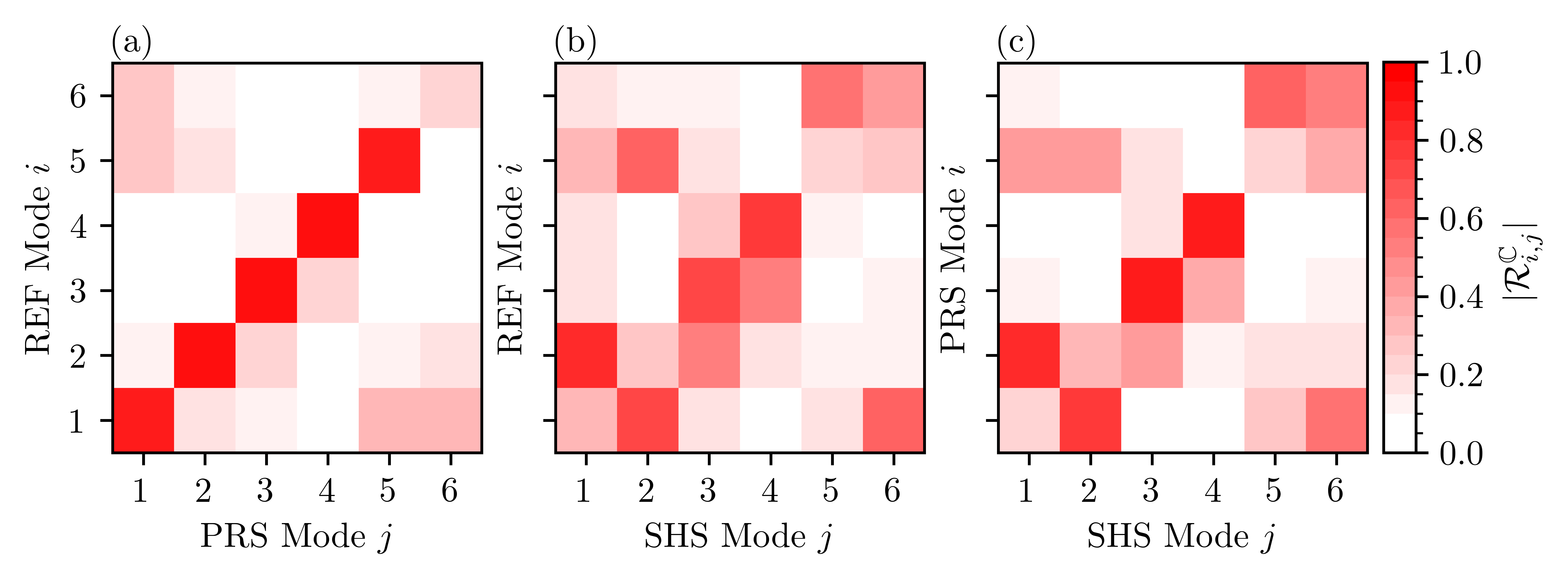}
  \caption{Correlation maps betwwen the modes of the (a) \ac{REF} and \ac{PRS} spheres, (b) \ac{REF} and \ac{SHS} spheres, and (c) \ac{PRS} and \ac{SHS} spheres.}\label{fig:Corr}
\end{figure*}

The similarity in the eigenvalues of the first five \ac{REF} and \ac{PRS} modes is reflected in the degree of diagonality of the cross-correlation matrix, as shown in figure~\ref{fig:Corr}a.
The cross-correlation between the \ac{SHS} and untreated spheres is less strongly diagonal, as shown in figures~\ref{fig:Corr}b and~\ref{fig:Corr}c, for the \ac{REF} and \ac{PRS} spheres, respectively. 
However, there are some consistent features between these cross-correlation maps which suggest that some \ac{SHS} modes are related to the modes of the untreated spheres.
The first and second \ac{SHS} modes best match the second and first \ac{REF} and \ac{PRS} modes, respectively.
The third and fourth modes of the \ac{SHS} are most highly correlated with the corresponding \ac{REF} and \ac{PRS} modes.
This is also consistent with the first modes of the \ac{REF} and \ac{PRS} spheres having a longer wavelength than the corresponding second modes, while the second \ac{SHS} mode has a longer wavelength than the first \ac{SHS} mode.
While the energy share of the first two \ac{SHS} modes is substantially lower than the leading modes of the untreated spheres, at approximately 15\% each compared with 20\% each, the first and second modes have comparable energy shares for each sphere.
The third and fourth modes have similar wavelengths and energy shares across all three spheres.


\begin{figure*}[h]
  \centering
  \includegraphics[width=6.4in]{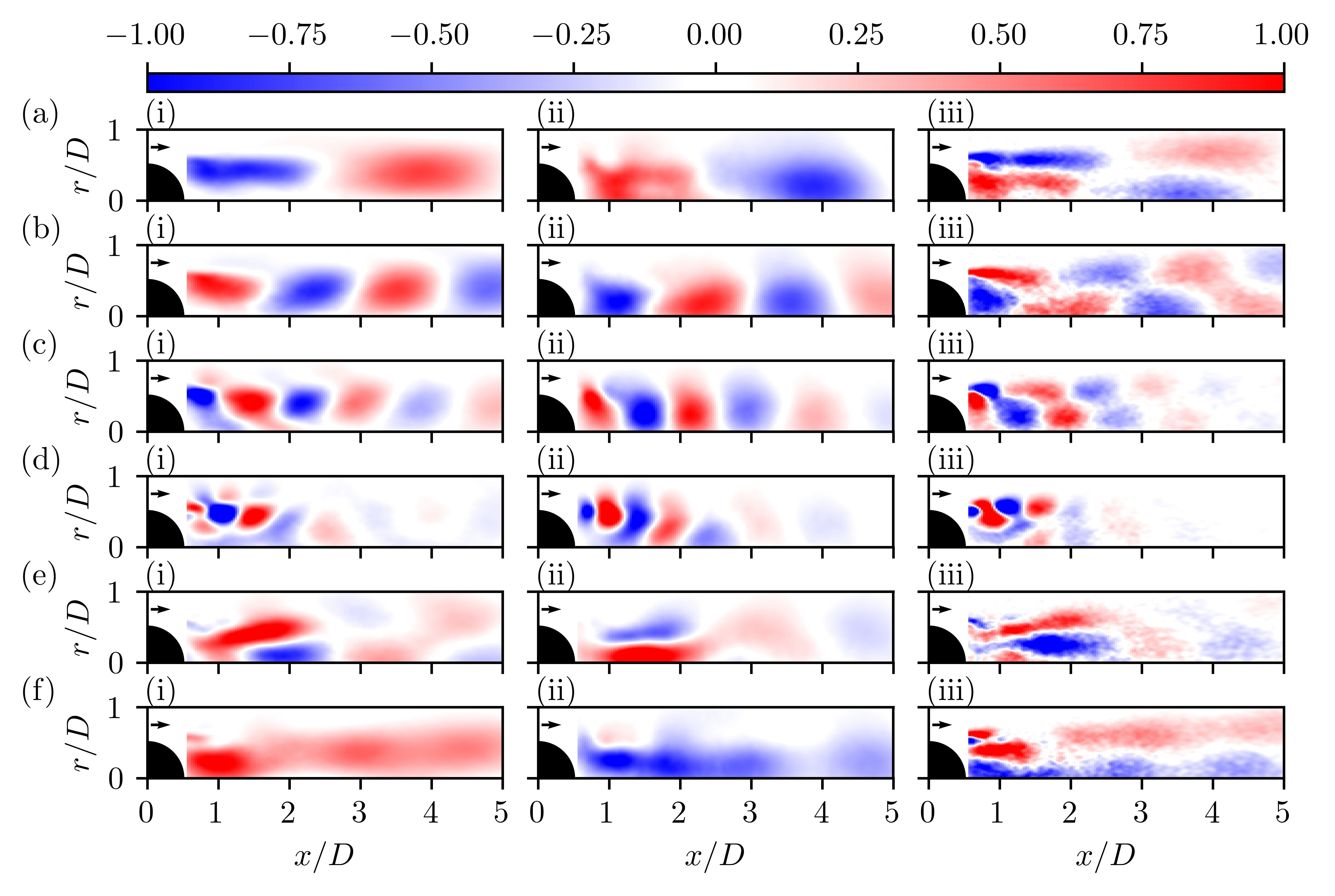}
  \caption{Normalised (i) streamwise and (ii) transverse velocity components, and corresponding (iii) out-of-plane vorticity for the six (a)-\@(f) \ac{DMD} modes of the \ac{REF} sphere.}\label{fig:REFmodes}
\end{figure*}

\begin{figure*}[h]
  \centering
  \includegraphics[width=6.4in]{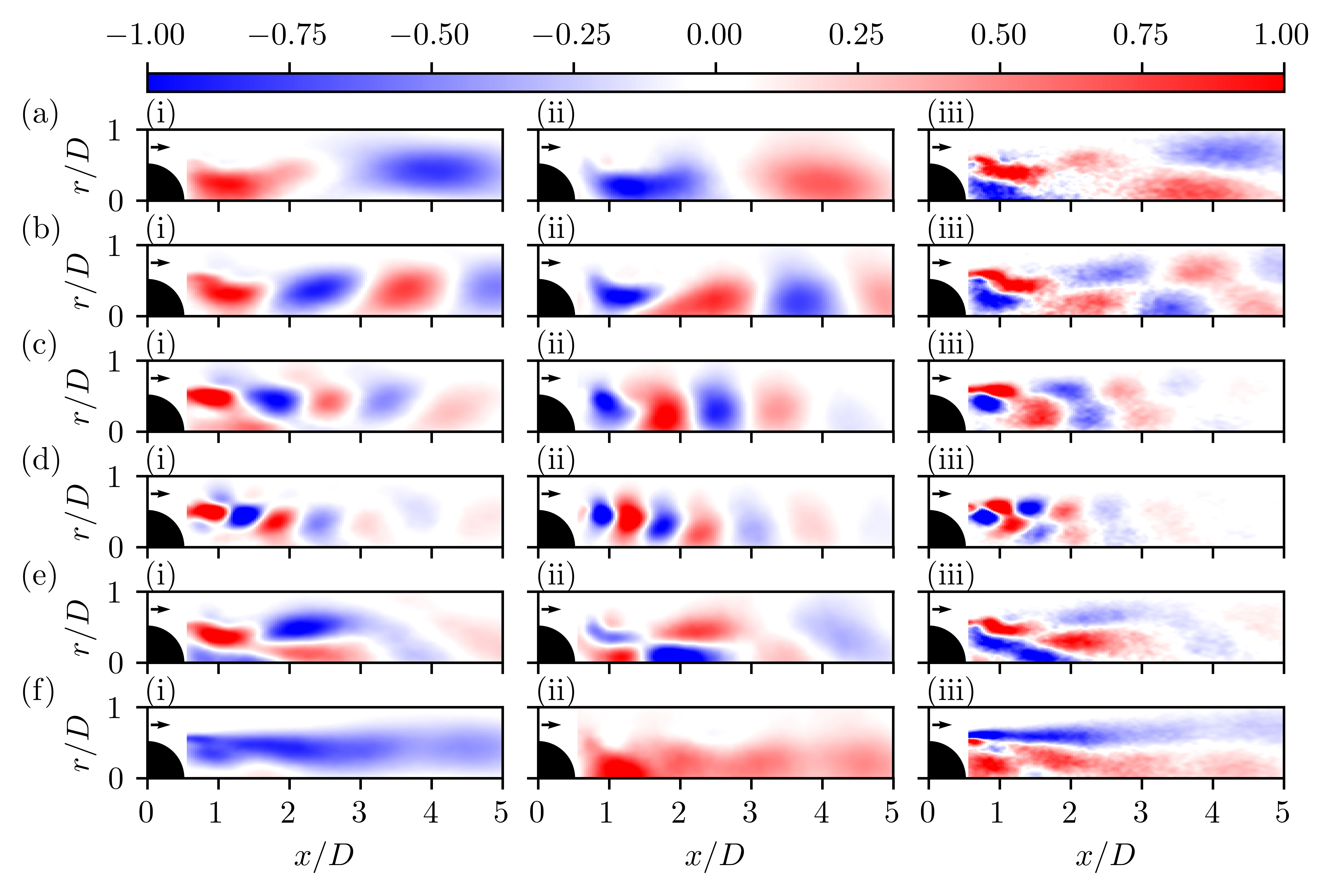}
  \caption{Normalised (i) streamwise and (ii) transverse velocity components, and corresponding (iii) out-of-plane vorticity for the six (a)-\@(f) \ac{DMD} modes of the \ac{PRS} sphere.}\label{fig:PRSmodes}
\end{figure*}

\begin{figure*}[h]
  \centering
  \includegraphics[width=6.4in]{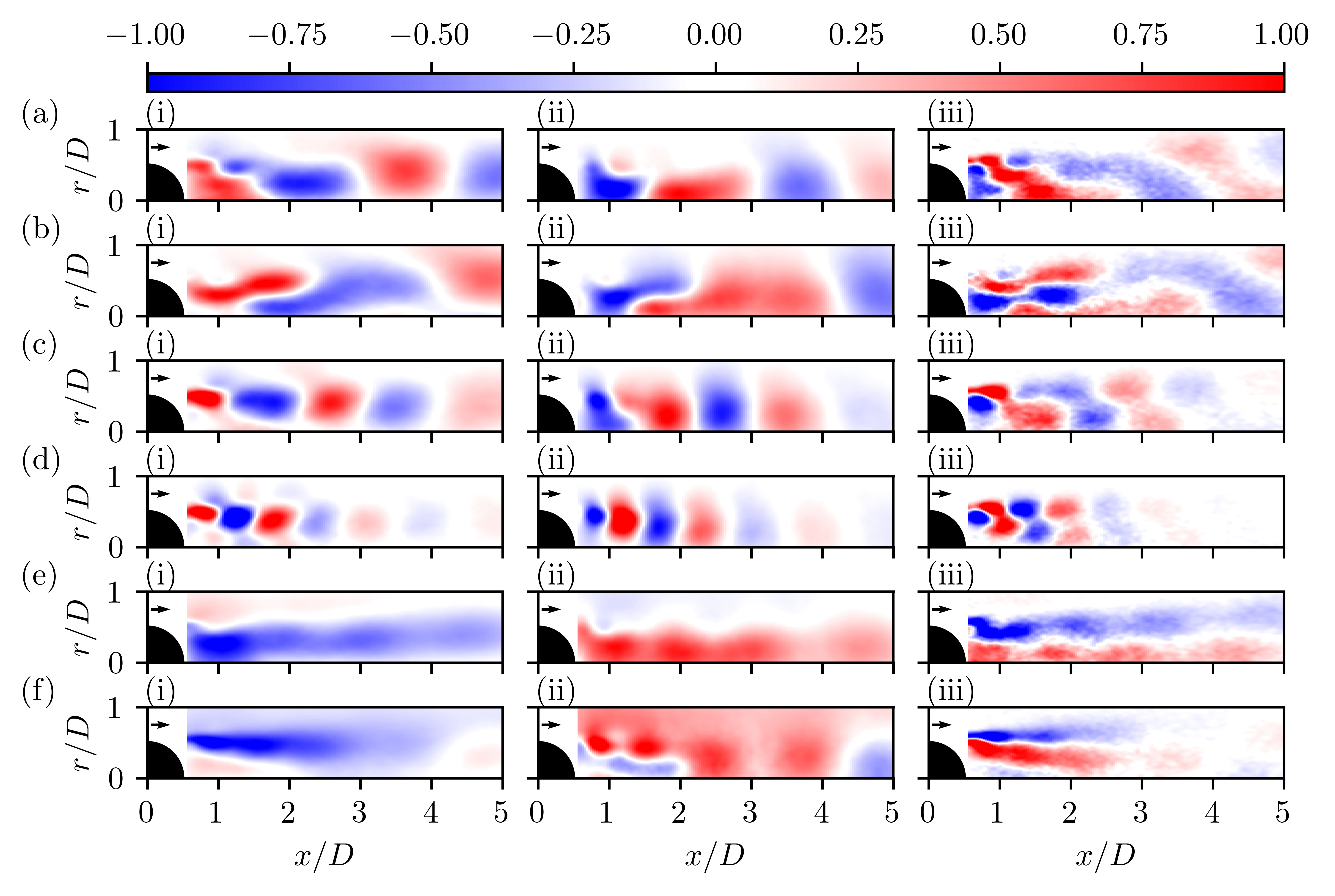}
  \caption{Normalised (i) streamwise and (ii) transverse velocity components, and corresponding (iii) out-of-plane vorticity for the six (a)-\@(f) \ac{DMD} modes of the \ac{SHS} sphere.}\label{fig:SHSmodes}
\end{figure*}

The first \ac{REF} mode has an energy share of $22.14$\%, a decay rate of $-0.02/D$, and a wavelength of $5.38D$.
The first \ac{PRS} mode has similar values, with an energy share of $21.03$\%, a decay rate of $-0.02/D$, and a wavelength of $5.54D$.
The first \ac{REF} mode is more strongly symmetric/antisymmetric than the \ac{PRS} mode, as shown by their $|\mathcal{R}_{+,-}^\mathbb{C}|$ values of 0.96 and 0.89, respectively.
The \ac{REF} mode, shown in figure~\ref{fig:REFmodes}a, features regions of streamwise and transverse fluctuations of alternating sign reminiscent of the first two \ac{POD} modes in~\cite{davey2025experimental} and first Hilbert \ac{POD} mode in~\cite{davey2026studying}, which represent a flapping in the wake.
The \ac{PRS} mode, shown in figure~\ref{fig:PRSmodes}a, features similar structures, although these are more distorted relative to the first \ac{POD} mode~\cite{davey2025experimental}.
The highest intensity in the streamwise component of the \ac{REF} mode, shown in figure~\ref{fig:REFmodes}a\@(i), remain centred at $r/D \approx 0.5$ for the entire measurement domain.
In contrast, the structures close to the sphere in the streamwise component of the \ac{PRS} mode, shown in figure~\ref{fig:PRSmodes}a\@(i), have the highest intensity closer to the centreline.
Further downstream, the structures in the streamwise component of the \ac{PRS} mode shift away from the centreline to match those of the \ac{REF} mode.
The shift in the structures in the \ac{PRS} mode relative to the \ac{REF} mode is reflected in the out-of-plane vorticity, which features longer horizontal structures on the \ac{REF} mode, as shown in figure~\ref{fig:REFmodes}a\@(iii), and shorter slanted structures in the \ac{PRS} case, shown in figure~\ref{fig:PRSmodes}a\@(iii).
From the eigenvalues, the closest match to these modes is the second mode of the \ac{SHS} sphere, which has an energy share of 14.82\%, a decay rate of $-0.03/D$, and a wavelength of $4.38D$.
In addition to having a shorter wavelength than the untreated spheres, the structure of this mode is distinct from that of the corresponding modes of the untreated spheres.
Directly behind the sphere, the structures in both the streamwise and transverse components, shown in figures~\ref{fig:SHSmodes}b\@(i) and~\ref{fig:SHSmodes}b\@(ii), respectively, curve towards the centreline before growing larger and shifting in the transverse direction further downstream.
While both of the untreated spheres feature structures which are completely separated in the streamwise direction, the structures of the \ac{SHS} mode overlap, with the downstream structure extending over the upstream structure around $x/D = 2$.
Further downstream, at $x/D = 4$, the structures no longer overlap, and the shape of the out-of-plane vorticity changes substantially from the approximately horizontal structures upstream, as shown in figure~\ref{fig:SHSmodes}b\@(iii).
 
The second \ac{REF} mode, shown in figure~\ref{fig:REFmodes}b, has an energy share of $19.88$\%, a decay rate of $-0.04/D$, and a wavelength of $2.64D$.
The second \ac{PRS} mode, shown in figure~\ref{fig:PRSmodes}b, has an energy share of $18.76$\%, a decay rate of $-0.03/D$, and a wavelength of $2.70D$.
The wavelengths of both modes are approximately half of the corresponding first modes, and their structures are similarly comparable.
This is consistent with the sixth and eighth \ac{POD} modes and third Hilbert \ac{POD} mode of \ac{REF} sphere in~\cite{davey2026studying}.
As in the case of the first mode, the \ac{REF} mode features horizontal structures, while the structures in the \ac{PRS} sphere are relatively deformed.
This is most evident in the out-of-plane vorticity, which is uniform in the streamwise direction in the \ac{REF} mode, as shown in figure~\ref{fig:REFmodes}b\@(iii), while the structures of the out-of-plane vorticity of the \ac{PRS} mode, shown in figure~\ref{fig:PRSmodes}b\@(iii), change in size and orientation as they advect downstream.
The first \ac{SHS} mode is the closest match to the second modes of the untreated spheres, with an energy share of 15.96\%, a decay rate of $-0.05/D$, and a wavelength of $2.99D$.
Unlike the second \ac{SHS} mode, which was closest to the first modes of the untreated spheres, this mode has a longer wavelength than the second modes of the untreated spheres.
While the downstream structures of the second \ac{SHS} mode are similar to the first \ac{SHS} mode, particularly in the out-of-plane vorticity, shown in figure~\ref{fig:SHSmodes}a\@(iii), the structure directly behind the sphere is significantly altered.
In addition to the large structures which alternate in sign downstream, smaller structures are present for $x/D < 2$.
In the streamwise component, shown in figure~\ref{fig:SHSmodes}a\@(i), there are two small regions of high intensity at $r/D = 0.5$ in addition to the larger structures present in the corresponding modes of the untreated spheres.
The corresponding region in the transverse component, shown in figure~\ref{fig:SHSmodes}a\@(ii), features a low-intensity positive region that is not present in the modes of the untreated spheres.
Additionally, the negative region closest to the sphere features a vertical protrusion on the upstream side, which is more intense than the corresponding structure in the second \ac{PRS} mode.
These structures are reflected in the out-of-plane vorticity of the first \ac{SHS} mode, shown in figure~\ref{fig:SHSmodes}a\@(iii), which features small staggered high-intensity regions close to the sphere and large diagonal structures further downstream, the latter of which are more reminiscent of the structures in the second \ac{SHS} mode.

The third modes of each sphere are closest matches to one another, with energy shares of $12.91\%$, $15.65\%$, and $13.56\%$, decay rates of $-0.11/D$, $-0.10/D$, and $-0.07/D$, and wavelengths of $1.58D$, $1.72D$, and $1.95D$, for the \ac{REF}, \ac{PRS}, and \ac{SHS} spheres, respectively.
All three modes, shown in figures~\ref{fig:REFmodes}c,~\ref{fig:PRSmodes}c and~\ref{fig:SHSmodes}c, respectively, feature regions of alternating sign in their streamwise component, which are located around $r/D = 0.5$ close to the sphere, and move closer to the centreline as they advect downstream.
The transverse components exhibit similar behaviour, although the these structures have an aspect ratio close to unity, while the structures in the streamwise velocity are longer in the streamwise direction, particularly close to the sphere.
The out-of-plane vorticity of these modes consists of uniform structures which fade in intensity as they travel downstream, which are most intense at $r/D=0.5$ and $x/D < 1$.
Unlike the first and second modes, there is no obvious change in the mode structures from either the addition of pores or superhydrophobic surface treatment.
Additionally, this mode does not clearly relate to the \ac{POD} modes in~\cite{davey2025experimental}, which were similar for the untreated spheres, but significantly altered in the \ac{SHS} case.
The fourth modes of each sphere are mutually best matches, with energy shares of $10.24\%$, $11.12\%$, and $10.03\%$, decay rates of $-0.19/D$, $-0.14/D$, and $-0.11/D$, and wavelengths of $0.99D$, $1.06D$, and $1.18D$, for the \ac{REF}, \ac{PRS}, and \ac{SHS} spheres, respectively.
These modes, shown in figures~\ref{fig:REFmodes}d,~\ref{fig:PRSmodes}d and~\ref{fig:SHSmodes}d, are similar in shape to the third modes of each sphere, but contain smaller structures corresponding to the shorter wavelengths of these modes.

The fifth modes of the \ac{REF} and \ac{PRS} spheres have similar energy shares of $10.24\%$ and $11.12\%$, respectively.
Their eigenvalues are also similar, with decay rates of $-0.09/D$ and $-0.07/D$, and wavelengths of $3.86D$ and $3.84D$, respectively.
The modes are similar in structure, although the \ac{PRS} mode, shown in figure~\ref{fig:PRSmodes}e, is phase-shifted relative to the \ac{REF} mode, shown in figure~\ref{fig:REFmodes}e.
The intensity of these modes drops off sharply after $x/D \approx 3$.
This corresponds to the structural change from parallel regions of opposing sign stacked in the radial direction to regions of alternating sign in the streamwise direction.

The sixth modes of both the \ac{REF} and \ac{PRS} spheres are characterised by wavelengths of $53.89D$ and $36.33D$, respectively, which are significantly longer than the measurement domain.
Despite this, these modes have the highest energy share for their respective spheres, at $23.17\%$ and $21.65\%$.
There is only a single structure in the velocity components of these spheres, as shown in figure~\ref{fig:REFmodes}f and~\ref{fig:PRSmodes}f, respectively.
However, the out-of-plane vorticity of these modes, and particularly the \ac{PRS} mode, resembles the structures in the first modes of the untreated spheres, but extends beyond the measurement domain.
The fifth and sixth \ac{SHS} modes, shown in figure~\ref{fig:SHSmodes}e and~\ref{fig:SHSmodes}f, have wavelengths of $65.38D$ and $10.44D$, respectively.
The fluctuations of intensity within regions of the same sign suggest these modes may represent a planar slice of out-of-plane structures.
Despite this, these modes have energy shares of $23.67\%$ and $21.96\%$, respectively, and thus represent almost half of the reduced-order dynamics of this sphere.

\section{Discussion}

The first four \ac{DMD} modes of each sphere, including the \ac{SHS} sphere, represent different instability scales in the shear layer of the wake.
This differs from the leading \ac{POD} modes in~\cite{davey2025experimental}, where the first six \ac{SHS} modes represented multiple permutations of the structure equivalent to the first two \ac{POD} modes of the untreated spheres. 
This was likely due to the interaction between the wake and the plastron resulting in a greater degree of three-dimensional variation in the wake, as observed in settling sphere experiments~\cite{castagna2018wake}.
Thus, the comparison of these modes allows for an examination of the effect of the superhydrophobic surface treatment on the shear layer instabilities.
The wavelengths of these modes are consistently longer for the \ac{PRS} case than the \ac{REF} case, and their decay rates are equivalent or slower.
Additionally, while the modes of the \ac{REF} sphere were consistently symmetrical, the symmetry was reduced in the \ac{PRS} case and further diminished in the \ac{SHS} case for some modes.
The structures near the sphere in the first \ac{PRS} mode are closer to the centreline than those in the first \ac{REF} mode, returning to a similar radial position further downstream.
This may be due to a tripping effect from the pores, similar to the effect of dimples on a golf ball~\cite{bearman1976golf}, delaying separation.
The effect of the superhydrophobic surface treatment on the shear layer is more varied.
The second \ac{SHS} mode has a significantly shorter wavelength and a faster decay rate than the corresponding modes of the untreated spheres.
The mode shape is also significantly altered compared to the untreated spheres, in which the structures were mostly horizontally aligned.
The \ac{SHS} mode instead features structures which overlap in the streamwise direction, with the structures extending downstream and away from the centreline to overlap with the portion of the upstream structures close to the centreline.
The first \ac{SHS} mode, which is the closest equivalent to the second modes of the untreated spheres, has a longer wavelength and faster decay rate.
The \ac{SHS} mode also features additional small structures above the regions which correspond to those present in the untreated sphere modes.
The larger structures in this mode are also closer to the centreline than those in the modes of the untreated spheres, which may result from the slip introduced by the surface treatment delaying flow separation.
The changes to the flow over the \ac{SHS} sphere are likely limited to the near-wake region, as the structures in these modes approach shapes similar to those of the modes of the untreated spheres towards the end of the measurement domain.
Without measurements at the surface of the sphere, which were not possible for this study due the laser reflections through the plastron, the exact cause of these changes cannot be determined.
However, the likely candidates are the introduction of slip delaying separation until further back on the sphere, the change in viscosity from the plastron dissipating or the injection of micro-bubbles into the wake.

The third and fourth modes of the \ac{PRS} sphere have longer wavelengths and decay more slowly than those of the \ac{REF} modes, with the \ac{SHS} modes having even longer wavelengths and slower decay rates than those of the \ac{PRS} sphere.
However, the change in the structure of these modes is less significant than those observed in the first and second modes.
The shapes of these modes are comparable across the three spheres, with the differences in wavelength being the only discernible variation.

The fifth modes of the untreated spheres vary significantly in shape from the first four modes, although they remain similar to each other.
As the structures in this mode extend through the centreline and change form around $x/D=3$, these modes may represent interactions between the recirculation region and the shear layers on either side of the sphere.
However, because the present study is focused on the shear layer and these modes clearly extend beyond the domain, the insight provided by these modes is limited. 

The sixth modes of the untreated spheres, as well as the fifth and sixth modes of the \ac{SHS} sphere, had wavelengths longer than could be resolved within the measurement domain.
Despite this, the mode shapes for the untreated spheres loosely resembled the structures in the leading four modes, but extended significantly in the streamwise direction.
These modes also had the largest energy share for their respective spheres, consistent with the energy of the other modes decreasing with higher wavenumbers.
In the \ac{SHS} case, both the fifth and sixth modes, which made up almost half of the energy share of the reduced-order dynamics, had wavelengths significantly longer than the domain.
This is similar to the shift mode represented by the \ac{DMD} of a transient cylinder wake presented in~\cite{noack2016recursive}, and may reflect the spatial decay of turbulence in the streamwise direction.
The presence of two of these modes for the \ac{SHS} case suggests that this decay is more complex than the untreated sphere cases, perhaps due to the interaction between the plastron and the flow.
The smaller structures visible in these modes may also suggest an effect on the turbulent fluctuations from the plastron dissolving into the flow at a regular rate.

Despite these limitations, the current analysis was able to elucidate modes representing the instabilities in the shear layer, enabling an assessment of the effect of the superhydrophobic surface treatment.
In contrast to the \ac{POD} analysis on the same spheres, the \ac{DMD} applied in the current work was able to identify changes in the flow due to the addition of pores.
The \ac{DMD} modes of the \ac{SHS} were not affected by the three-dimensionality of the dominant instability in the wake, which dominated the leading \ac{POD} modes in~\cite{davey2025experimental}.
Thus, the leading modes of the \ac{SHS} sphere were able to clearly capture the equivalent instabilities to the untreated spheres.
Additionally, the fifth and sixth modes of the \ac{SHS} sphere captured smaller structures within their main structure, which were not present in the equivalent modes of the untreated spheres.

\section{Concluding Remarks}

The streamwise and transverse velocities in the near wake of spheres with and without superhydrophobic surface treatment were measured using \ac{2C-2D} \ac{MCCD}-\ac{PIV}.
The mean planar velocities were then determined and used to calculate the instantaneous velocity fluctuations, which were separated into domains above and below the sphere's centreline, centred on the shear layer.
\ac{DMD} was performed on the fluctuations in each domain, using a shift in the streamwise direction as an analogue for a temporal separation.
The \ac{DMD} modes in each domain were compared and adjusted for phase shifts before averaging.
The resulting \ac{DMD} modes were sorted by energy share, with long wavelengths assigned lower rankings.
The complex cross-correlation was used to identify equivalent modes across spheres to assess the effect of pores and superhydrophobic surface treatment on the shear-layer dynamics.
The leading four \ac{DMD} modes of each sphere represent equivalent instabilities, with the wavelengths, decay rates and mode shapes altered by the superhydrophobic surface treatment.
In addition to these modes, the untreated spheres featured similar fifth and sixth modes, with the sixth mode having a wavelength significantly longer than the measurement domain despite having the highest energy share.
Similarly, the fifth and sixth modes of the \ac{SHS} sphere had wavelengths significantly longer than the measurement domain, with long structures featuring spatially regular increases in intensity.
These smaller structures may be the result of periodic changes in the decaying turbulence due to air dissolving into the fluid as the plastron dissipates.
Where the leading \ac{POD} modes of the \ac{SHS} sphere were contaminated by the increased three‑dimensionality of the dominant untreated‑sphere structures, the current application of \ac{DMD}, using spatial advection as an analogue for time, was able to identify the primary instabilities in the wake’s shear layer.
While the addition of pores to the sphere's surface had some effect on these instabilities, the superhydrophobic surface treatment produced more substantial changes.
Additionally, while all three spheres included long-wavelength modes with high energy shares, the \ac{SHS} sphere featured small, regularly spaced structures within these modes.

\FloatBarrier{}
\section*{Acknowledgements} 

The authors acknowledge the support of the Australian Research Council (ARC) in funding this research through Linkage Infrastructure, Equipment and Facilities (LIEF) and Discovery grants.
Shaun Davey gratefully acknowledges the support of the Australian Commonwealth Government through a Research Training Program (RTP) Scholarship.
This research was undertaken with the support of resources provided via a Monash HPC-NCI Merit Allocation from the National Computational Infrastructure (NCI Australia), an NCRIS-enabled capability supported by the Australian Government.

\bibliographystyle{ieeetr}
\footnotesize\bibliography{refs} 

\end{document}